\documentclass[10pt, conference, letterpaper]{IEEEtran}
\IEEEoverridecommandlockouts
\usepackage{cite}
\usepackage{amsmath,amssymb,amsfonts}
\usepackage{algorithmic}
\usepackage{graphicx}
\usepackage{textcomp}
\usepackage{xcolor}
\usepackage{paralist}
\usepackage{multirow}
\usepackage{multicol}
\usepackage[para]{threeparttable}
\def\BibTeX{{\rm B\kern-.05em{\sc i\kern-.025em b}\kern-.08em
    T\kern-.1667em\lower.7ex\hbox{E}\kern-.125emX}}
\begin{document}

\title{Enjoy the Untrusted Cloud: A Secure, Scalable and Efficient SQL-like Query Framework for \\Outsourcing Data}

\author{Yaxing~Chen,
	Qinghua~Zheng,~\IEEEmembership{Member,~IEEE,}
	Dan~Liu,
	Zheng~Yan,~\IEEEmembership{Senior Member,~IEEE,}\\
	Wenhai~Sun,~\IEEEmembership{Member,~IEEE,}
	Ning~Zhang,~\IEEEmembership{Member,~IEEE,}
	Wenjing~Lou,~\IEEEmembership{Fellow,~IEEE,} \\
	and~Y. Thomas~Hou,~\IEEEmembership{Fellow,~IEEE}
	\thanks{Y. Chen and Q. Zheng are with the School of Electronic and Information Engineering, Xi'an Jiaotong University, Xi'an, Shaanxi, China. Email: cyx.xjtu@gmail.com, qhzheng@mail.xjtu.edu.cn}
	\thanks{D. Liu and Z. Yan are with the State Key Lab on Integrated Services Networks, School of Cyber Engineering, Xidian University, Xi'an, Shaanxi, China. Email: 15596173220@163.com, zyan@xidian.edu.cn; Z. Yan is also with the Department of Communications and Networking, Aalto University, Espoo, Finland. Email: zheng.yan@aalto.fi}
	\thanks{W. Sun, is with the Department of Computer and Information Technology, Purdue University, West Lafayette, Indiana, USA. Email: whsun@purdue.edu}
	\thanks{N. Zhang, is with the Department of Computer Science and Engineering, Washington University in St. Louis, St. Louis, Missouri, USA. Email: zhang.ning@wustl.edu}
	\thanks{W. Lou and Y. T. Hou are with the Department of Computer Science, Electrical and Computer Engineering, respectively, Virginia Tech, Blacksburg, Virginia, USA. Email: \{wjlou, thou\}@vt.edu}}

\maketitle

\begin{abstract}
While the security of cloud remains a concern, a common practice is to encrypt data before outsourcing them for utilization. One key challenging issue is how to efficiently perform queries over the ciphertext. Conventional crypto-based solutions, e.g. partially/fully homomorphic encryption and searchable encryption, suffer from low performance, poor expressiveness and weak compatibility. An alternative method that utilizes hardware-assisted trusted execution environment, i.e., Intel SGX, has emerged recently. On one hand, such work lacks of supporting scalable access control over multiple data users. On the other hand, existing solutions are subjected to the key revocation problem and knowledge extractor vulnerability. In this work, we leverage the newly hardware-assisted methodology and propose a secure, scalable and efficient SQL-like query framework named QShield. Building upon Intel SGX, QShield can guarantee the confidentiality and integrity of sensitive data when being processed on an untrusted cloud platform. Moreover, we present a novel lightweight secret sharing method to enable multi-user access control in QShield, while tackling the key revocation problem. Furthermore, with an additional trust proof mechanism, QShield guarantees the correctness of queries and significantly alleviates the possibility to build knowledge extractor. We implemented a prototype for QShield and show that QShield incurs minimum performance cost.
\end{abstract}


\section{Introduction}
The cloud computing paradigm, characterized by \textit{convenience}, \textit{elasticity} and \textit{low-cost}, demonstrates a great success in the past decade \cite{CC}. Organizations typically need to deploy their application services (ASes) to remote servers that are not in charge by themselves, which leaves these organizations no choice but to trust the cloud in outsourcing their data for utilization. This security assumption that the cloud is fully trusted, however, is not always valid, especially when outsourced data contain sensitive information such as medical records and personal identifiable information, since the cloud may suffer from the malfunctions or even compromise of the AS as well as system software, e.g. OS and VM hypervisor. Hence, many cloud-based systems depend on cryptography to protect confidential data when being transmitted, computed and/or stored. One of biggest challenges is how to efficiently compute over encrypted data while not hindering data utilization, such as queries that are performed frequently in many systems \cite{C_Issue, cryptdb, sc_streamforce, sc_polystream}. Typically, a generic query can be denoted by a SQL-like expression and interpreted as a query plan, i.e. directed acyclic graph (DAG) of computational operators for execution, such as \textit{projection}, \textit{selection}, \textit{aggregation}, \textit{union} and \textit{join}.

Partially/Fully homomorphic encryption (PHE/FHE) \cite{Ding_PHE, FHE} is a fundamental technology to solve the problem in the literature. However, such pure crypto-based solutions at present suffer from severe expressiveness and performance issues \cite{Hermetic, SGX_SA_power, ZeroTrace}. For example, FHE introduces many orders of magnitude overheads and cannot handle computational tasks of a practical scale. Searchable Encryption (SE) enables search over encrypted data. Albeit there exist extensive investigations along this research line, current SE implementation is still not satisfactory \cite{rw_sun_rearguard}. Notably, building upon various crypto primitives, different SE schemes support different search types, e.g., single keyword, multi-keywords and range, with different index structures, and are not compatible with each other. Besides, SE focuses on conditional information retrieval, which can only be viewed as a \textit{selection} operator. As such, existing schemes inherently cannot be applied or extended to support generic (full-featured) queries.

Recognizing the possible hardware-assisted trusted execution environment (TEE), i.e., Intel SGX, is used in \cite{rw_sun_rearguard, rw_fuhry, Opaque} as an alternative promising countermeasure, but presents several limitations. Specifically, Hardidx in \cite{rw_fuhry} only considers a single data user scenario; it does not support access control over multiple data users, which is a fundamental functionality for utilization of data outsourced to the cloud. Rearguard in \cite{rw_sun_rearguard} improved that but has scalability problem, since it requires each authorized data user to perform cumbersome remote attestation for authentication and verification. In contrast with the two systems who focus on information retrieval, Opaque \cite{Opaque} adopts a generic SQL-like query model. Its primary goal, however, also does not take multi-user access control into account. Moreover, the above works overlook the fact that a TEE can still be influenced by an untrusted host AS - even though the integrity and correctness of a TEE can be attested, the expected invocation of TEE interfaces cannot be guaranteed - this will potentially cause some covert, non-trivial security and privacy issues, especially after the entrusted key materials are delivered to a TEE. For instance, the key revocation problem, firstly identified by \cite{RMF}, means that an untrusted host AS may selectively drop network packets such that it is problematic for a valid remote entity to timely and effectively revoke crypto keys within a TEE. Another example is knowledge extractor \cite{SGX_RANE}, which indicates that the global internal state of a TEE can be influenced by an untrusted host AS by not following the expected protocol workflow and thus open door for exposing confidential information. Please refer to \cite{SGX_SA_cache, Interface} for concrete real world cases. Furthermore, a practical design for a SGX-based query system is implementing each computational operator as a unique TEE interface. This incurs another challenging issue, that is, how to guarantee the integrity of the whole distributed workflow per query \cite{VC3}.

In this paper, we propose a secure, scalable and efficient query framework called QShield to enable flexible utilization of outsourced data; it adopts Intel SGX to establish hardware-assisted TEEs (also called enclaves) in the untrusted cloud platform to protect the confidentiality and integrity of sensitive data run inside. By considering a more generic SQL-like query model, QShield is capable of handling majority of common computational tasks. Notably, we define four operators in QShield, i.e., \textit{projection}, \textit{selection}, \textit{aggregation}, \textit{join}, and each of them is implemented as a unique enclave functional interface. We remark that an enclave theoretically can realize arbitrary computational operators. In addition, we exploit the widely-adopted, flexible document-oriented data model in QShield to enable compatibility since cloud applications in web, mobile, social or IoT scenarios often use different data models, such as relational tables, key-value items and data streams \cite{Data_Models_1, Data_Models_2}. 

With regard to supporting access control over multiple data users, a straightforward method, adopted by previous work, is as follows: Let the data owner securely provision the crypto key and his/her access policy to the enclave through remote attestation; Then, the enclave loads encrypted data, recovers authorized data for a specific user, and queries over them. Such a method, however, exposes fatal flaws since only TEEs in the cloud are trustworthy. First, it is a risky thing to hand out the crypto key to an enclave in the long run; as mentioned above, a compromised host AS can either make use of the enclave performing malicious computation or extract the crypto key from it. Albeit Chen et al. in \cite{RMF} proposed a novel "heartbeat" synchronization protocol to enable key revocation on demand, their work is vulnerable to network failure and requires a trusted broker to be always online \cite{SRM}. Second, each authorized data user still needs to perform remote attestation to verify the integrity and correctness of the enclave, which is a cumbersome process and thus causes poor scalability. In order to tackle these issues, we present a novel, lightweight secret sharing method. The core idea is let the data owner assign an attested enclave a secret share $sk_{a}$ and each authorized data user another unique secret share $sk_{b}^{i}$. The crypto key $sk$ can be reconstructed and used for decryption if and only if the $sk_{b}^{i}$ is delivered to the enclave that holds $sk_{a}$ per query. Once the authorized data are recovered, the crypto key $sk$ is erased. As such, neither data users nor the enclave have full capability to recover the encrypted data with their own secret shares outside the scope of current query; the non-trivial key revocation problem can then be avoided. Another beneficial gain is that the enclave and a data user can effectively authenticate with each other through their unique secret shares rather than cumbersome remote attestation.

Nevertheless, due to the fact that Intel SGX provides no functionality to dynamically attest distributed computational workflow, we need a mechanism to enforce that the recovered data within enclave are only used to serve the current query. Otherwise, a compromised host AS may execute arbitrary enclave operators other than desired ones, generating incorrect results; Worse still, it may build knowledge extractor to expose sensitive information. In QShield, we propose a two-part solution to solve this problem. First, we leverage an endurance indicator denoted by $\omega$ to restrict the times that the recovered data as well as its derived intermediate results can be computed by enclave operators. Second, we view the query process as a finite state machine (FSM) and make the enclave record and output execution footprints of state transition as a workflow proof for auditing.

Our key contributions are summarized as follows.
\begin{inparaenum}
	\item Building upon the off-the-shelf hardware-assisted TEE, i.e., Intel SGX, we propose a practical secure query framework named QShield for outsourcing data in the untrusted cloud. Compared to existing crypto-based solutions, the proposed framework is efficient and powerful.
	\item By supporting common SQL-like query expressions and flexible document-oriented data model, QShield can be easily adopted by most of cloud-based query application scenarios.
	\item Under threats caused by the limitation of SGX architecture, we present a secure, lightweight secret sharing method to make QShield capable of realizing scalable multi-user access control. 
	\item We also propose a trust proof mechanism to ensure correctness of queries as expectation with auditing and meanwhile greatly alleviating the possibility to build knowledge extractor.
\end{inparaenum}


\section{Background}\label{P}

\subsection{Intel SGX}
Intel Software Guard Extensions (SGX) is a promising hardware-assisted trusted computing technology. It provides \textit{memory isolation} \cite{SGX_isolation}, which enables a host application set up a protected execution environment, called enclave, such that code and data run inside are resilient to attacks from privilege software, including OS kernel and VM hypervisor. Function calls between the untrusted host application and enclave are through well-designed ECALL/OCALL interfaces. Specifically, a call to enclave is referred to as an ECALL and OCALL allows enclave codes to call untrusted functions outside. Such an architecture implies that the invocation to ECALLs is unreliable since it is still under control of the untrusted host application. Intel SGX also offers two auxiliary functionalities: \textit{remote attestation} and \textit{storage sealing} \cite{SGX_attestation_seal}. The former makes a distant entity capable of verifying the authenticity of an enclave, checking the integrity of desired code running inside and meanwhile establishing a secure communication channel with the enclave. The latter allows to store enclave data in untrusted storage outside for future recovery, in case of server shutdown, system failure, and/or power outage. Please refer to \cite{SGX_explained} for a more thorough technical analysis about Intel SGX. 

\noindent\textbf{Limitations.} Intel SGX, however, is reported to suffer from various vulnerabilities caused by either physical or digital attacks. Among which, one mainstream methodology is exploiting side-channels, including cache timing, power analysis, branch shadowing, and the most recently discovered foreshadow transient execution, etc. to expose confidential information \cite{SGX_SA_cache, SGX_SA_power, SGX_SA_branch, SGX_SA_foreshadow}. Moreover, the fore-mentioned implication with regard to ECALL invocation can incur potential security threats, for example, the key revocation problem \cite{RMF}. Besides the listed side-channels, it is also found that utilizing ECALL invocation can build knowledge extractor \cite{Code-reuse}. Furthermore, a malicious OS can launch DoS attack to disrupt operations of enclave functions due to the fact that it is still in charge of the underlying enclave resource allocation.

\subsection{Bilinear Maps}
We briefly review a few facts about groups with efficiently computable bilinear maps\cite{Bilinear_Map}, based on which we implement the proposed secret sharing mechanism in QShield.

Let $\mathbb{G}_{1}$ and $\mathbb{G}_{2}$ be two multiplicative cyclic groups of prime order $p$, and $g$ be a generator of $\mathbb{G}_{1}$. An efficiently computable bilinear map $e: \mathbb{G}_{1} \times \mathbb{G}_{1} \rightarrow \mathbb{G}_{2}$ defined over the two groups satisfies the following properties:
\begin{enumerate}
	\item Bilinearity: for all $a$, $b$ $\in \mathbb{Z}_{p}$, there exists $e(g^{a}, g^{b}) = e(g, g)^{ab}$
	\item Non-degeneracy: $e(g, g) \neq 1$
	\item Computability: for any $u$, $v$ $\in \mathbb{G}_{1}$, there exists an efficient algorithm to compute $e(u, v)$.
\end{enumerate}

\noindent\textbf{Decisional Bilinear Diffie-Hellman Assumption.} The security of our secret sharing mechanism is based on the Decisional BDH assumption. Basically, let $a$, $b$, $c$, $z$ be chosen randomly from $\mathbb{Z}_{p}$, there exists no probabilistic polynomial-time algorithm $\mathcal{B}$ that can distinguish the tuple $(A = g^{a}, B = g^{b}, C = g^{c}, e(g, g)^{abc})$ from the tuple $(A = g^{a}, B = g^{b}, C = g^{c}, e(g, g)^{z})$ with more than a negligible advantage $\epsilon$, that is,
\begin{equation*}
\begin{aligned}
|Pr[\mathcal{B}(A, B, C, & e(g,g)^{abc}) = 0]  \\
& - Pr[\mathcal{B}(A, B, C, e(g,g)^{z}) = 0]| \leq \epsilon,
\end{aligned}
\end{equation*}
where the probability is computed over the randomly chosen generator $g$, the randomly chosen $a$, $b$, $c$, $z$ in $\mathbb{Z}_{p}$, and the random bits consumed by $\mathcal{B}$.

\section{Models and Assumptions}\label{MA}
\subsection{System Model}
\begin{figure}[!t]
	\centering
	\includegraphics[width=3.5in]{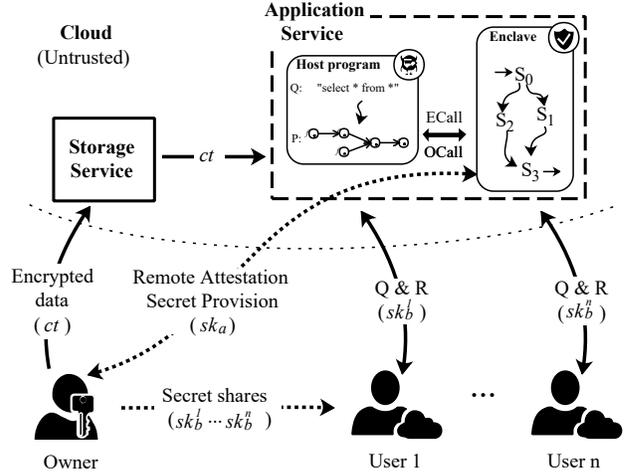}
	\caption{System Model}
	\label{system_model}
\end{figure}

As illustrated in Fig.~\ref{system_model}, QShield involves one data owner and multiple data users. The data owner uses a block cipher encryption algorithm such as AES[GCM] making use of a symmetric key $sk$ to protect his/her sensitive data outsourced to the cloud. He/She also defines access permissions for each data user in the system. Upon receiving a query request from an authorized data user, the enclaved cloud AS can retrieve those ciphertexts and compute over them for response.

More concretely, the AS first creates a dedicated enclave for the data owner. Provided that it is successfully attested, the data owner can establish a secure communication channel with the enclave, through which a secret share $sk_{a}$ of $sk$ and an access policy $pol$ towards involved data users are delivered. Meanwhile, the data owner assigns each authorized data user the other unique secret share $sk_{b}^{i}$ of $sk$. Here, we assume that the communication process between the data owner and data users is secure and happens out-of-band. Given $sk_{b}^{i}$, a data user now is capable of building valid query requests to the AS. In QShield, a request is defined to include a SQL-like query expression and a cryptographically protected token. On the cloud side, upon receiving a query request, the AS first forwards the token to the enclave, enabling the enclave itself create an initial state for the query, and meanwhile transforms the query expression into a query plan. Then, the AS invokes corresponding operators for computation accordingly. As the whole computation per query globally runs to completion, the enclave will go through several intermediate states and reach to a final state, i.e., results. In parallel with this state transition process, the enclave records footprints of execution to establish a trust proof for the query. At last, the enclave cryptographically encapsulates the results and returns the ciphertext along with a self-signed trust proof to the AS for query response.

\subsection{Data Model}\label{DM}
We opt for a document-oriented data model in QShield due to its flexibility in being compatible with most types of data models, e.g., relational tables, key-value items and data streams, exploited by current web, mobile, social, as well as IoT applications. Specifically, QShield employs the notion of \textit{document} $D$ as a basic logical unit for data storage and query, which, encoded in JSON, contains a set of \textit{attributes} $\{A_{1}, \cdots, A_{n} | A_{j} :=\ <name_{j}, value_{j}>, j = 1, \cdots, n\}$, where $name_{j}$ is a description of the \textit{attribute} $A_{j}$ and $value_{j}$ is a value for that attribute. It also uses the notion of \textit{collection} $C$ to represent a group of \textit{documents}, each of which includes a same set of attribute descriptions. When outsourced to the cloud, a \textit{collection} of size $r$ can be divided into multiple parts, each stored as a data \textit{file} $F$ of size $s$. Formally, $C = \{F_{1}, \cdots, F_{m-1}, F_{m} | F_{k} = \{D_{(k-1)*s + 1}, D_{(k-1)*s + 2}, \cdots, D_{(k-1)*s + s}\}, k = 1, \cdots, m-1; F_{m} = \{D_{(m-1)*s+1}, \cdots, D_{(m-1)*s+(r \% s)}\}\}$.

What follows is the formal definition of four computational operators supported by QShield.

Projection operator ($\pi$): It traverses a \textit{collection} $C$ and iteratively prunes unselected \textit{attributes} ($\notin \{A_{p}, \cdots, A_{q}\}$) from each involved \textit{document}. Algebraically, $C' \leftarrow \pi(A_{p}, \cdots, A_{q})(C)$.
	
Selection operator ($\sigma$): It bases a predicate $P$ over \textit{attributes}, e.g., $C.A_{j} = b$, to filter out the \textit{documents} not satisfying $P$. Algebraically, $C' \leftarrow \sigma(P)(C)$.
	
Aggregation operator ($\phi$): It applies aggregate functions $f$, e.g. $average$, over an attribute $A_{j}$ of documents in a \textit{collection} $C$. Algebraically, $value \leftarrow \phi(f, A_{j})(C)$
	
Join operator ($\gamma$): It bases a predicate $P$, e.g., $C_{1}.A_{p} = C_{2}.A_{q}$, to perform natural join over two \textit{collections} $C_{1}$ and $C_{2}$. Algebraically, $C' \leftarrow \gamma(P)(C_{1}, C_{2})$.

\subsection{Threat Model}\label{TM}
The main goal of QShield is exploiting Intel SGX to build a system that can efficiently and correctly answer queries over outsourced data in multi-user settings while offering strong privacy guarantee to the data owner. We assume that the queries themselves (expressions) are not sensitive - only their answers and input datasets are - and that data users will honestly execute protocol but also desire to query data beyond their own permissions (honest-but-curious). Note that, QShield consists of multiple enclave functional interfaces; once they are loaded in the cloud platform, the data owner is capable of attesting their integrity and correctness. 

Besides the standard SGX threat model where an attacker may control the cloud's software stack, including hypervisor and OS, we consider a more powerful attacker who may also compromise the AS that hosts the enclave. As such, 
\begin{inparaenum}
	\item the attacker may prevent the data owner from revoking the previously entrusted crypto key on demand by selectively dropping network packets;
	\item the attacker may trigger black-swan level information leakage, once she (painstakingly but potentially) builds knowledge extractor to recover the long-term crypto key within the enclave;
	\item the attacker may arbitrarily invoke enclave functional interfaces, resulting in incorrect results and contributing to knowledge extractor;
\end{inparaenum} 

Denial-of-service, enclave bugs and interface interruption, physically measuring and manipulating SGX-enabled CPU package, and collusion between data users with the cloud are outside our scope for now.

\section{QShield Construction}\label{F}
\subsection{Countermeasures}
We introduce two novel methods to make QShield against attacks highlighted in Section~\ref{TM}, so as to accomplish its promising design goals.

\paragraph{Secret Sharing}
We remark that the fact a long-term crypto key is handed out to an enclave significantly contributes to those attacks. First, such a full-capability (in terms of ciphertext decryption) enclave may be maliciously used by the host AS. Second, the enclave may not be as impregnable as a wall of iron in a long run; the crypto key can be exposed. Third, the entrusted crypto key may fail to be revoked as the data owner desires. This seems a dilemma for the data owner, since he/she needs to entrust the crypto key to the enclave for data utilization. We propose a novel idea to solve this. In a nutshell, we partition the full decryption capability among involved participants, i.e., the enclave and data users, and it can be restored only when the enclave handles queries issued by a data user. On one hand, neither the enclave nor the data user has the full capability to recover ciphertexts outside the scope of a query. Risks regarding to exposure of a long-term enclave crypto key are also eliminated. On the other hand, it inherently supports authentication between the enclave and data users, thus avoiding cumbersome remote attestation between them. 

In our context, such secret sharing scheme can be defined over a tree-based access structure. Specifically, the root node of the access tree is an AND gate: its left child represents the enclave; its right child is an OR gate with $n$ children, each representing a data user. We note that this construction has been widely used in cryptography, like attribute-based encryption. Similarly, we base bilinear maps design a crypto primitive called $\mathcal{E}$ to facilitate secret sharing in QShield, which consists of following algorithms:

\leftmargini=0mm
\begin{itemize}
	\item \noindent$sk, sk_{a}, \{sk_{b}^{1}, sk_{b}^{2}, \cdots, sk_{b}^{n}\} \leftarrow setup(1^{\lambda}, n)$: This algorithm takes as inputs a security parameter $\lambda$ and a non-zero positive integer $n$. It outputs following secret components: a symmetric secret key $sk$, a secret share $sk_{a}$, and $n$ secret shares $sk_{b}^{i}\ (1 \leq i \leq n)$. More specifically, it first defines a bilinear group $\mathbb{G}_{1}$ of prime order $p$ with a generator $g$ and a bilinear map $e : \mathbb{G}_{1} \times \mathbb{G}_{1} \rightarrow \mathbb{G}_{2}$ that has properties of bilinearity, non-degeneracy and computability. Then, it randomly picks up two numbers $r$ and $m$ from $\mathbb{Z}_{p}$ and for each $i, 1 \leq i \leq n$, it uniformly chooses a number $t_{i}$ at random from $\mathbb{Z}_{p}$. At last, it computes $sk = \mathcal{H}(e(g, g)^{m})$, $sk_{a} = \{g^{t_{1}}, \cdots, g^{t_{n}}, e(g, g)^{(r + m)}\}$, $sk_{b}^{i} = g^{\frac{2r+m}{t_{i}}}\ (1 \leq i \leq n)$. 
	\item \noindent$ct \leftarrow encrypt(sk, msg)$: This algorithm takes in the symmetric secret key $sk$ and a message $msg$. It then encrypts the data using symmetric encryption, i.e., $ct \leftarrow E.Enc(sk, msg)$ and finally outputs the corresponding ciphertext $ct$.
	\item \noindent$msg^{i} \leftarrow decrypt(pol, sk_{a}, sk_{b}^{i}, ct)$: This algorithm takes as inputs an access policy $pol$, the secret share $sk_{a}$, the secret share $sk_{b}^{i}$, and the ciphertext $ct$. It outputs the corresponding data $msg^{i}$ that can be accessed by the one who owns $sk_{b}^{i}$. More specifically, the algorithm first looks up the $sk_{a}$ to retrieve $e(g,g)^{(r+m)}$ and $g^{t_{i}}$, and computes $e(g^{t_{i}}, sk_{b}^{i}) = e(g^{t_{i}}, g^{\frac{2r+m}{t_{i}}}) = e(g,g)^{(2r+m)}$. Then, it reconstructs the secret key $sk$ by computing $\frac{(e(g,g)^{(r+m)})^{2}}{e(g,g)^{(2r+m)}} = e(g,g)^{m}$ and performing $\mathcal{H}(e(g,g)^{m})$. Finally, it decrypts the cyphertext $ct$ with $sk$, i.e., $pt \leftarrow E.Dec(sk, ct)$ and obtains $msg^{i}$ by filtering out those unauthorized data in $pt$ according to the access policy $pol$. By design, this algorithm is executed by an enclave interface, thus the confidentiality of data and the unique workflow of this algorithm can be enforced.
\end{itemize}

\paragraph{Trust Proof}
Albeit the novel secret sharing idea blocks out most of mentioned attacks, it is still possible for a compromised host AS to maliciously call enclave operators for computation over the recovered native message $msg^{i}$. Obviously, this can incur incorrect results. There also exists a covert threat, i.e., building knowledge extractor to expose sensitive information. For example, computation over $msg^{i}$ under different invocation sequences often leads to different results. An attacker may utilize such difference along with prior knowledge to discover sensitive information. In order to alleviate these risks, on one hand, we leverage an endurance indicator denoted by $\omega$ to restrict the times that $msg^{i}$ and its derived intermediate messages can be accessed by enclave operators; on the other hand, we record the execution trace of each enclave operator per query and chain them together as a trust proof. By design, once the final result is obtained, the $\omega$ will be set as $0$, such that no access over relevant messages are allowed any more. Suppose that the compromised host AS invokes some (even one) enclave operator(s) outside the scope of current query, it will not be able to construct a valid trust proof for auditing, since there is no enough $\omega$ allowing to do that. Such a countermeasure also greatly reduces training datasets for knowledge extractor construction.

\subsection{Framework Description}
Next, we present a description of QShield protocols, which include System Setup, Data Upload, Data Query, and Policy Update. Note that, $E$ represents a $256$-bits authenticated symmetric key encryption scheme, which consists of $Enc(\cdot)$ and $Dec(\cdot)$ algorithms. $PKE$ represents a $256$-bits indistinguishability security under chosen plaintext attack (IND-CPA) public key encryption scheme, which consists of $KeyGen(\cdot)$, $Enc(\cdot)$ and $Dec(\cdot)$ algorithms. $S$ stands for a $256$-bits existentially unforgeable signature scheme, which consists of $KeyGen(\cdot)$, $Sign(\cdot)$ and $Verify(\cdot)$ algorithms. $\mathcal{H}(\cdot)$ denotes a $256$-bits collision resistant hash function. $f_{*}$ stands for enclave computational operators $*$, where $*$ can be a projector $\pi$, a selector $\sigma$, an aggregator $\phi$, or a joiner $\gamma$.

\textbf{System Setup} The data owner first selects a security parameter $\lambda$ and defines the scale of data users $n$, i.e., the maximum number of data users allowed in the system. Then, he/she invokes $\mathcal{E}.setup(1^{\lambda}, n)$ to generate all secret components $sk$, $sk_{a}$, $\{sk_{b}^{1}, sk_{b}^{2}, \cdots, sk_{b}^{n}\}$. Besides, the data owner creates an access policy $pol :=\ <(uid, cids)_{j}>_{j=0}^{n-1}$, where the $uid$ and $cids$ of an entry $j$ represents the ID of a data user and all IDs (initialized as NULL) of \textit{collections} authorized to the data user, respectively. On the cloud side, the AS creates a dedicated enclave $\mathbb{E}_{App}$ for the data owner. It subsequently performs an ECALL to let the enclave generate a $256$-bits public key pair, i.e., $(pk_{e, msg}, sk_{e, msg}) \leftarrow PKE.KeyGen(1^{\lambda})$, and a $256$-bits signature key pair, i.e., $(vk_{e, sign}, sk_{e, sign}) \leftarrow S.KeyGen(1^{\lambda})$.  The $pk_{e, msg}$ and $vk_{e, sign}$ are output as public system parameters. Next, the data owner verifies that the $\mathbb{E}_{App}$ is correctly deployed and executed on a genuine Intel SGX-enabled CPU platform through remote attestation. During this process, he/she also establishes a secure communication channel with $\mathbb{E}_{App}$ by negotiating a symmetric secret key $sk_{comm}$. After that, the data owner makes use of $sk_{comm}$ to escort the secret share $sk_{a}$ to $\mathbb{E}_{App}$, i.e., $ct_{sk_{a}} \leftarrow E.Enc(sk_{comm}, sk_{a})$ by the data owner and $sk_{a} \leftarrow E.Dec(sk_{comm}, ct_{sk_{a}})$ by the enclave. For each data user $i\ ( 1 \leq i \leq n)$ registered in the system, the data owner selects a unique secret share $sk_{b}^{i}$ and securely delivers it to the specific data user.

\textbf{Data Upload} When a new \textit{document} $msg$ is ready for outsourcing, the data owner obtains its ciphertext $ct_{msg}$ by executing $\mathcal{E}.encrypt(sk, msg)$ and performs $\mathcal{H}(ct_{msg})$ to create a unique ID $did$ for the \textit{document}. Provided that the \textit{document} belongs to an existing \textit{collection}, the data owner just uploads $\{did, ct_{msg}\}$ to the cloud and inserts it to the \textit{collection}. Otherwise, the data owner needs to create a new \textit{collection} with ID $cid$ in the cloud, defines access permissions over the \textit{collection} for data users, and iteratively updates each authorized data user's \textit{collection} ID list $cids$ in the policy $pol$ with $cid$. The updated $pol$ will then trigger the execution of Policy Update protocol described below. At last, the data owner uploads $\{did, ct_{msg}\}$ to the newly generated \textit{collection} in the cloud.

\textit{Remarks: the exploited one \textit{document} per upload paradigm can benefit a real-time cloud application. Beyond that, the data owner can improve network throughput by buffering a certain amount of \textit{documents} before performing an upload action.} 

\textbf{Data Query} In this protocol, an authorized data user $i$ requests queries towards the \textit{collections} shared by the data owner. A query $req$ by design consists of two parts: a SQL-like expression $q$ and a unique query token $tk$, which is generated by encrypting the data user's secret share $sk_{b}^{i}$, an endurance indicator $\omega$, and a monotonically increasing positive number $c$ with the enclave's public key $pk_{e, msg}$, i.e, $tk \leftarrow PKE.Enc(pk_{e, msg}, (sk_{b}^{i}, \omega, c))$. Note that, $\omega$ is calculated based on the current query expression.

Upon the AS receives a query request $req$ from the data user, it will perform the following two steps. The first one, called \textit{unlock}, enables the enclave $\mathbb{E}_{App}$ to recover the \textit{collections} allowed to be accessed by the data user. Specifically, the AS first retrieves relevant encrypted \textit{collections} $ct$ and forwards them along with the token $tk$ to $\mathbb{E}_{App}$, within which the enclave executes $PKE.Dec(sk_{e, msg}, tk) \rightarrow \{sk_{b}^{i}, c, \omega\}$. Then, $\mathbb{E}_{App}$ checks whether or not the current request has been previously handled by comparing $c$ with a $c'$ (initialized as $-1$). If $c$ is smaller than $c'$, the enclave will deny to serve the current request; otherwise, it updates $c'$ with $c$ and further calls $\mathcal{E}.decrypt(pol, sk_{a}, sk_{b}^{i}, ct)$ to recover authorized \textit{collections} $msg^{i}$. 

\begin{figure}[!t]
	\centering
	\includegraphics[width=3.4in]{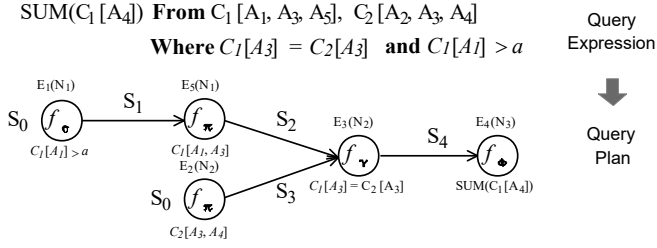}
	\caption{A Running Example for Query Transformation}
	\label{query_process}
\end{figure}

The other step is called \textit{query}, which invokes enclave computational operators to perform computation over the $msg^{i}$ and delivers a response $res$ to the data user. In this step, the AS first retrieves query expression $q$ from the request $req$ and transforms it into a query plan $\tau$, which by design is representable by a directed acyclic graph (DAG). To be specific, vertices in the DAG are de facto well-designed enclave computational operators, i.e. $f_{\pi}$, $f_{\sigma}$, $f_{\phi}$, $f_{\gamma}$, and directed edges illustrate the data flow between them. Figure~\ref{query_process} shows a running example where the data user issues a query over two \textit{collections} $C_{1}[A_{1}, A_{3}, A_{5}]$ and $C_{2}[A_{2}, A_{3}, A_{4}]$. The query first filters out from $C_{1}$ the \textit{documents} that the value of $A_{1}$ is less equal than $a$. Then, it performs projection of $C_{1}$ and $C_{2}$ on $A_{3}$ and $[A_{3}, A_{4}]$, respectively, and further joins the two \textit{collections} under $C_{1}[A_{3}] = C_{2}[A_{3}]$. At the end, the query computes the sum of $C_{2}[A_{4}]$.

\begin{figure}[!t]
	\centering
	\includegraphics[width=3.4in]{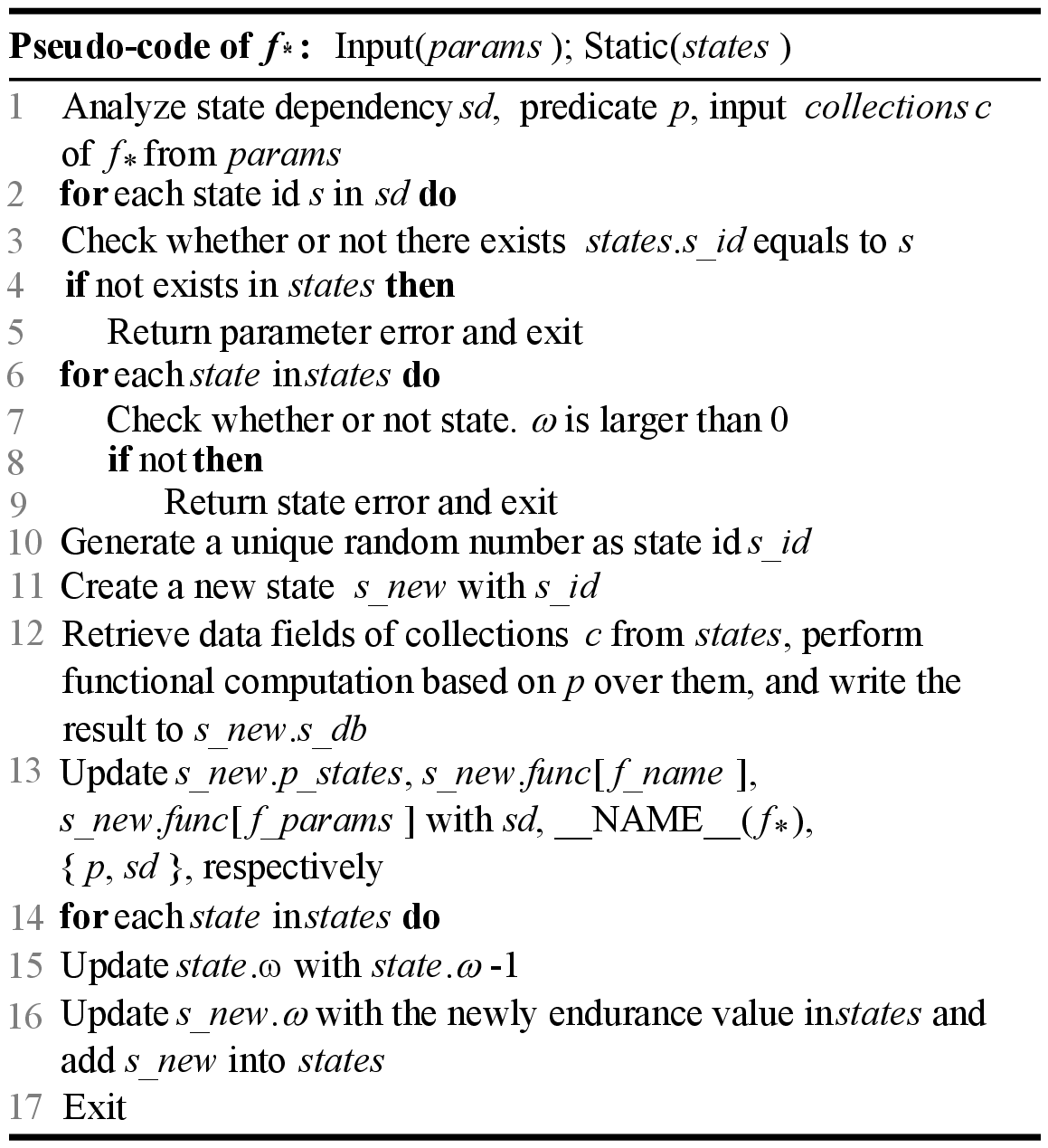}
	\caption{Pseudo-code of the Enclave Computational Operator $f_{*}$}
	\label{pseudocode}
\end{figure}

Afterwards, the AS begins to schedule corresponding enclave operators $f_{*}$ in accordance with the query plan $\tau$. As this process proceeds, the enclave $\mathbb{E}_{App}$ will go through several states $S_{0}, S_{1}, \cdots, S_{k}$. Here, a state is defined as a struct that includes $s\_id$, $p\_states$, $func$, $s\_db$ and $w$, where $s\_id$ is the state ID; $p\_states$ records all IDs of previous states that derives $S$ under the function of $f_{*}$; $func$ records metadata of $f_{*}$, which includes its name $f\_name$ and its parameters $f\_params$; $s\_db$ stores the real workload of state data; $w$ records the maximum times that the current state can be accessed by enclave operators; Notably, $S_{0}$ is initialized with the recovered native message $msg^{i}$ and the endurance indicator $\omega$ in the token. Figure~\ref{pseudocode} demonstrates the pseudo-code of $f_{*}$.

After the enclave reaches to the final state $S_{k}$, it can construct a response for the current query. More concretely, the response includes two parts: \textit{result} and \textit{trust proof}. The \textit{result} is generated by encrypting the data field $s\_db$ in $S_{k}$ with the data user's secret share $sk_{b}^{i}$, formally, \textit{result} $\leftarrow E.Enc(sk_{b}^{i}, S_{k}.s\_db)$, and the \textit{trust proof} is a tuple $(tp, \sigma_{tp})$, where $tp$ is execution trace obtained from all states $S_{0}, S_{1}, \cdots, S_{k}$ and $\sigma_{tp}$ is a signature signed by the enclave, i.e., $\sigma_{tp} \leftarrow S.Sign(sk_{e, sign}, tp)$. When the data user gets the response, he/she can recover the result by computing $E.Dec(sk_{b}^{i}, result)$ and audit it with the \textit{trust proof} $tp$.

\textbf{Policy Update} This protocol handles all events of policy update, i.e., user add/remove and access permission modification. When a new data user $j\ (1 \leq j \leq n)$ joins in the system, the data owner first selects a unique secret share $sk_{b}^{j}$ and securely sends it to the data user. Then, he/she defines a set of \textit{collections} that can be accessed by the data user and computes a unique ID $uid$ for the data user by hashing $sk_{b}^{j}$, i.e., $uid \leftarrow \mathcal{H}(sk_{b}^{j})$. At last, the data owner updates the policy $pol$ in $\mathbb{E}_{App}$ with a new item $(uid, cids)$ through their previously established secure channel. As for the remaining two circumstances, the data owner only needs to either delete or alter the corresponding item in the $pol$ for the specific user. 

\textit{Remarks: one potential issue here is that the update commands may not be received by an enclave. This can be solved by just requiring a response for each update from the enclave. Supposing it is caused by network failure, the data owner can continue to send the update command until he/she receives a response. Provided that the host AS is compromised, the potential corresponding damage is limited and affordable before the data owner takes some remedial measures, since the secret sharing mechanism guarantees that the enclave has no full capability to decrypt ciphertexts.}

\subsection{Distributed Mode}
With data becoming available in larger quantities and system requiring higher throughputs, a common practice for the AS is exploiting distributed computation, where tasks are processed across multiple computational nodes in the cloud. We remark that QShield can be extended so as to support such context. To this end, we introduce the notions of \textit{broker} and \textit{worker} in QShield: a \textit{broker} is a dedicated enclave created for the data owner; a \textit{worker} is a general enclave serving all data users, which implements one of computational operators $f_{\pi}$, $f_{\sigma}$, $f_{\phi}$, $f_{\gamma}$. As illustrated in Fig.~\ref{query_process}, we have $E_{1}$(N\textsubscript{1}), $E_{2}$(N\textsubscript{2}), $E_{3}$(N\textsubscript{2}), $E_{4}$(N\textsubscript{3}), $E_{5}$(N\textsubscript{1}), where $E_{x}$(N\textsubscript{y}) represents the $x$\textsuperscript{th} \textit{worker} in the $y$\textsuperscript{th} node. Compared with the stand-alone mode, protocols in QShield have following changes. 

Other than the \textit{query} step in Data Query protocol, interactions with the enclave in all original protocols are replaced by with the \textit{broker}. We also modify the following two protocols accordingly.

\noindent\textbf{System Setup:} On behalf of the data owner, the \textit{broker} validates the intactness of codes of \textit{workers} and the credibility of their hosting SGX-enabled platforms through remote attestation, and meanwhile builds interconnected secure channels with all \textit{workers} by negotiating a common communication key.

\noindent\textbf{Data Query:} In the \textit{query} step, after the \textit{broker} obtains $msg_{i}$ and $\omega$, it creates an initial state $S_{0}$ with the two items for the current query. Then, based on the query plan $\tau$, the AS successively schedules \textit{workers} instead for computation. Notably, when a \textit{worker} finishes its computational task, it generates a new state $S_{j}$ $(j = 1, \cdots, k)$ and informs the \textit{broker} to record its execution trace. Besides, the last \textit{worker} will forward the final state, i.e., results, to the \textit{broker} for response construction. All communications between the \textit{broker} and \textit{workers} during above process are through the previously established secure channels.

\section{Implementation}\label{I}
We implemented a prototype for QShield in C using the Intel SGX SDK (v$2.4$) for linux. It is tested on a SGX-enabled platform that runs an Intel Kaby Lake i$7$-$7700$ processor at $3.60$GHz with $16$ GiB of RAM and Ubuntu $16.04$ operating system. The code is compiled using gcc in the debug mode.

\textbf{Library Migration.}
In order to offer strong security guarantees, Intel SGX sets two restrictions on enclave development: 
\begin{inparaenum}
	\item an enclave only supports portion of legal CPU instructions; and
	\item libraries used by an enclave must be statically linked. 
\end{inparaenum}
Therefore, we have to port the adopted third-party libraries for use in the SGX development environment. By analysis, we need migrate the Stanford pairing-based cryptography (PBC) library and its dependency GNU multiple precision (GMP) library into the prototype. The lucky thing is that Intel already demonstrates how to transform GMP into an enclave-safe trusted library. Following the same procedure, we can do PBC migration. However, the native PBC library does not decouple I/Oes from its core functionality, so we need to rewrite the affected functions.

The native PBC library relies on a fundamental data struct called $field\_s$, which defines three I/O function pointers, i.e., $out\_str$, $snprint$ and $out\_info$. These interfaces output elements of a field in a human-readable manner and have different instances when the field is initialized using different algorithms. While operating on a non-enclave-safe C File* type, their native implementations are at odds with the SGX development environment. On the other hand, OCALL functions are typically implemented by an untrusted host AS; it is weird to assemble OCALLs into an enclave-safe library. As such, we work out a method for an enclave to collect messages from its linked libraries, built upon which the three I/O interfaces in the PBC library are re-implemented to facilitate the output of human-readable information. In a nutshell, the enclave-safe library first caches output messages in its global memory stack and then makes it accessible through public interfaces. To be specific, we implement three auxiliary interfaces in the ported PBC library, i.e., $sgx\_init\_msg(\cdot)$, $sgx\_clear\_msg(\cdot)$, and $sgx\_get\_msg(\cdot)$, where $sgx\_init\_msg(\cdot)$ initializes a global memory region for recording output message while $sgx\_clear\_msg(\cdot)$ frees the memory region. $sgx\_get\_msg(\cdot)$, used between the two interfaces, is capable of reading the memory region.

\section{Evaluation}\label{E}
\subsection{Correctness and Security}
\textbf{Correctness.} QShield is correct, if for the data owner and all authorized data users in the system, the following three statements hold:
\begin{inparaenum}
	\item the probability that a query issued by a data user can compute over unauthorized data is negligible;
	\item the probability that other involved entities including the cloud platform and host AS can obtain sensitive data is negligible;
	\item the probability that a data user accepts an incorrect result is negligible.
\end{inparaenum}

Built upon the presented secret sharing and trust proof mechanisms, such correctness is realized by follows. A valid query by design brings a unique cryptographical token, which can only be utilized by the enclave that holds secret share $sk_{a}$. Once obtaining $sk_{b}^{i}$, the enclave can recover data from loaded ciphertexts by calling  $\mathcal{E}.decrypt(\cdot)$. During this process, the enclave also enforces access control policy on behalf of the data owner, eliminating unauthorized data accesses. Thanks to the security properties of SGX, the cloud platform and host AS cannot spy on the recovered data. Besides, should a compromised host AS arbitrarily invokes enclave operators, the incorrect result will be detected by the end user through its associated trust proof.

\textbf{Security.} Next, we discuss the security properties achieved by QShield.
\paragraph{Confidentiality of crypto key and outsourced data} This property is guaranteed by both the adopted enclave interfaces and the $\mathcal{E}$ scheme.

At System Setup, the data owner generates all secret components $sk$, $sk_{a}$, $sk_{b}^{i} (i = 1, \cdots, n)$ (by $\mathcal{E}.keygen(\cdot)$) and establishes a secure communication channel with the enclave through remote attestation, under which the $sk_{a}$ is escorted to the enclave. The data owner also assigns $sk_{b}^{i}$ to each authorized data user securely. At Data Upload, the data owner encrypts his/her data with $sk$ (by $\mathcal{E}.encrypt(\cdot)$) and outsources the ciphertexts $ct$ to the cloud for utilization. At the moment, even the enclave and data users both have secret shares, no one can decrypt the ciphertexts. At Data Query, a valid query request brings the crypto-protected $sk_{b}^{i}$ to the enclave that holds $sk_{a}$ for $sk$ reconstruction. Once the enclave recovers authorized data $msg^{i}$, the crypto key $sk$ will be erased (by $\mathcal{E}.decrypt(\cdot)$). The subsequent computations over $msg^{i}$ are performed by enclave operators and the final results are encrypted before delivering to the user. Due to the fact that the integrity and correctness of enclave interfaces are guaranteed by SGX, the confidentiality of crypto key and outsourced data are realized as long as the $\mathcal{E}$ scheme is secure.

The $\mathcal{E}$ scheme in essentials consists of two independent crypto components: secret sharing scheme $\mathcal{E}_{A}$, which generates a symmetric key $sk$ and its shares $sk_{a}, sk_{b}^{i}$; and symmetric key encryption scheme $\mathcal{E}_{B}$, which encrypts and decrypts data with $sk$. In QShield, $\mathcal{E}_{B}$ is IND-CPA secure since we adopt an authenticated symmetric encryption scheme, e.g., AES[GCM]. Therefore, $\mathcal{E}$ is secure if and only if $sk$ can be securely reconstructed by authorized parties and the IND-CPA security of data encrypted with $sk$ cannot be compromised. As described above, the security of $sk_{a}$ transmission and $sk$ reconstruction is ensured by SGX, so we just need to prove that data users possessing $sk_{b}^{i}$ cannot recover $sk$ without the knowledge of $sk_{a}$, even if they are colluded, and vice versa. We recall that our secret sharing scheme makes use of widely-adopted bilinear maps implementing a tree-based access structure. It has been proven secure under the standard bilinear Diffie-Hellman (BDH) assumption in many crypto works \cite{Goyal_abe, Souza_pvss}. Therefore, without $sk_{a}$, any combination of $sk_{b}^{i}$ cannot constitute a valid authorized set for the access structure, and vice versa, and thus the blind factor of the root node for $sk$ cannot be recovered unless the BDH assumption is compromised.

\paragraph{Trustworthy query processing} This property is guaranteed by both the remote attestation of SGX and the trust proof mechanism. During remote attestation, the enclave will generate a cryptographic tag called \textit{quote} for the program running inside the enclave, with which the data owner can verify the integrity and correctness of critical enclave interfaces that take sensitive data as input, i.e., $f_{\pi}$, $f_{\sigma}$, $f_{\phi}$, $f_{\gamma}$. With regard to the correctness of distributed workflow per query, the constructed trust proof guarantees that the host AS cannot arbitrarily invoke enclave operators for computation. 

\paragraph{Scalable multi-user access control} QShield exploits a straightforward access control list (ACL) to realize authorization to data users. Each data user is only associated with an item $(uid, cids)$ in the policy $pol$. When a data user queries over the outsourced data, the enclave on behalf of the data owner will check the $pol$ and decide whether or not the current query is valid. When the policy needs to be updated, it only requires to modify corresponding items in $pol$. With regard to authentication, the data owner in QShield assigns each data user a secret share $sk_{b}^{i}$ and an enclave the other secret share $sk_{a}$. While pairing the two shares for crypto key reconstruction, the enclave and the data user can also authenticate with each other. As the above process produces much less temporal and spatial overhead, QShield achieves scalable access control over multiple users.

\newcommand{\tabincell}[2]{\begin{tabular}{@{}#1@{}}#2\end{tabular}} 
\begin{table*}[!t]
	\caption{Comparison with Other Similar Works}
	\centering
	\label{work_diff}
	\begin{threeparttable}
		\fontsize{7.5}{12.0}\selectfont 
		\begin{tabular}{c|c|c|c|c|c|c|c|c|c|c|c|c|c|c}
			\multicolumn{1}{c|}{\multirow{3}{*}{\bfseries Work}} & \multicolumn{2}{c|}{\multirow{2}{*}{\tabincell{c}{\bfseries Computing\\\bfseries Paradigm}}} & \multicolumn{3}{c}{\multirow{2}{*}{\bfseries Access Control}} &  \multicolumn{9}{|c}{\bfseries Trustworthy Data Processing}\\\cline{7-15}
			\multicolumn{1}{c|}{} & \multicolumn{2}{c|}{} & \multicolumn{3}{c|}{} &\multicolumn{6}{c|}{\bfseries Operators (op.)\tnote{1}} & \multicolumn{2}{c|}{\bfseries Method} & \multirow{2}{*}{\tabincell{c}{\bfseries Order-of-Magnitude\\ \bfseries (op. exe. time)}} \\\cline{2-14}
			\multicolumn{1}{c|}{} & Batch & Real-time & Model\tnote{2} & Granularity & Privilege\tnote{3} & $\pi$ & $\sigma$ & $\phi$ & $\gamma$ & $\zeta$ & I/D & crypto & TEE &\multicolumn{1}{|c}{}\\
			\hline
			\bfseries CryptDB\textsuperscript{\cite{cryptdb}} & $\surd$ & $\bigcirc$ & $N/A$ & \textit{coarse} & \textit{r/w} & $\surd$ & $\surd$ & $\surd$\textsubscript{\textit{sum}} & $\surd$ & $\surd$ & $\surd$ & $\surd$ & $\bigcirc$ & $ms$\\
			\hline
			\bfseries Streamforce\textsuperscript{\cite{sc_streamforce}} & $\bigcirc$ & $\surd$ & ACL & \textit{coarse} & \textit{queries} & $\surd$ & $\surd$ & $\surd$\textsubscript{\textit{sum}} & $\surd$ & $\surd$ & $\bigcirc$ & $\surd$ & $\bigcirc$ & $s$\\
			\hline
			\bfseries PloyStream\textsuperscript{\cite{sc_polystream}} & $\bigcirc$ & $\surd$ & ABAC  & \textit{fine} & \textit{op. set} & $\surd$ & $\surd$ & $\surd$\textsubscript{\textit{sum}} & $\surd$ & $\bigcirc$ & $\bigcirc$ & $\surd$ & $\bigcirc$ & $s$\\
			\hline
			\bfseries QShield & $\surd$ & $\surd$ & ACL &  \textit{coarse} & \textit{r/w} & $\surd$ & $\surd$ & $\surd$ & $\surd$ & $\surd$ & $\surd$ & $\bigcirc$ & $\surd$ & $ms$\\
		\end{tabular}
		\begin{tablenotes}
			\footnotesize
			\item[1] $\pi$ for \textit{selection} or \textit{filter}; $\sigma$ for \textit{projection} or \textit{map}; $\phi$ for \textit{aggregation}; $\gamma$ for \textit{join}; $\zeta$ for \textit{range}; I/D for \textit{insert} or \textit{delete}; $\surd$\textsubscript{\textit{sum}} for only summation supported.
			
			\item[2] \textit{N/A} for not applicable; ACL for access control list; ABAC for attribute-based access control.
			
			\item[3] \textit{r/w} denotes that a data user can perform arbitrary read and write operations on authorized data; \textit{queries} demotes that a data user can only perform specific queries on authorized data; \textit{op.set} refers to that a data user can perform specific operators on authorized data.
			
		\end{tablenotes}
	\end{threeparttable}
\end{table*}

\subsection{Performance}
We evaluate QShield in terms of computational overhead of two core components, that is, enclave operators and the $\mathcal{E}$ scheme.

\begin{figure}[!t]
	\centering
	\includegraphics[width=3.5in]{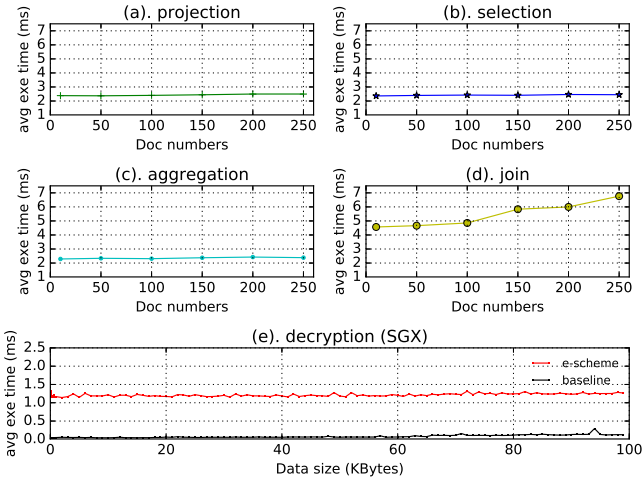}
	\caption{Performance evaluation of implemented operators and $\mathcal{E}$-scheme with Intel SGX}
	\label{eval}
\end{figure}

\paragraph{Enclave operators} 
Figure~\ref{eval} (a) - (d) illustrates the performance of implemented enclave operators $f_{\pi}$, $f_{\sigma}$, $f_{\phi}$, and $f_{\gamma}$, respectively (without optimization). It can be observed that these operators are very efficient with millisecond ($ms$) order of magnitude. However, the performance degradation of $f_{\gamma}$ is more fierce than the other three ones as the number of \textit{documents} grows. This is because the join operation needs nested iterations through two whole collections to search all solutions. 

The experimental results also demonstrate that our method exploiting hardware-assist TEE shows great competitiveness to the pure crypto solutions like partially/fully homomorphic encryption (FHE/PHE). In a recent work by Ding. et al. \cite{Ding_PHE}, they implemented several low-level computational operators based on Paillier's PHE, e.g., \textit{addition}/\textit{subtraction}, \textit{multiplication}, \textit{absolute}, \textit{comparison} and \textit{equality test}. Their experimental analysis shows that a single \textit{comparison} operation requires $8$ module exponentiation and the corresponding execution time on a PC with similar configuration than ours is about $50 ms$ under a weak security level (i.e., the security parameter $n$ is set as $1024$ bits). When considering a stronger security level (e.g., $n = 2048$ bits), it will reach to approximate $375 ms$. The execution time for a single string \textit{comparison} operation in our method, however, is much smaller ($\mu s$ order of magnitude). Due to the fact that each of the four enclave operators consists of numerous string comparisons, it is unrealistic to implement them using PHE.

\paragraph{$\mathcal{E}$ scheme} 
This crypto primitive is executed by both the data owner and the enclave in QShield. We focus on analyzing the performance of $\mathcal{E}.decrypt(\cdot)$ with Intel SGX, in that $\mathcal{E}.keygen(\cdot)$ is an one-time constant operation at system initialization, and $\mathcal{E}.encrypt(\cdot)$ just computes a straightforward AES ciphertext. 

In our setting, an enclave executes $\mathcal{E}.decrypt(\cdot)$ with different input sizes, ranging from $1$ Byte to $100$K Bytes. This simulates the real world where data to be processed varies greatly. For example, a batch-based query needs to perform computation over historical data with thousands of documents, while a real-time-based query requires to compute over few documents (even one) every time to satisfy timeliness. Figure~\ref{eval} (e) shows the performance of decryption of implemented $\mathcal{E}$ scheme (without optimization). We use a native AES decryption realized by Intel SGX as baseline, which, by comparison, lacks of crypto key reconstruction. It can be concluded:
\begin{inparaenum}
	\item the decryption performance of $\mathcal{E}$ scheme is efficient with millisecond ($ms$) order of magnitude and has slow degradation as the size of input grows;
	\item the main computational overhead of $\mathcal{E}.decrypt(\cdot)$ is caused by the time-consuming pairing operation when reconstructing the crypto key. 
\end{inparaenum}

\subsection{Comparison with Similar Systems}

Table~\ref{work_diff} compares QShield with other systems \cite{cryptdb, sc_streamforce, sc_polystream} that support SQL-like queries and meanwhile achieve similar security design goals, i.e., multi-user access control and trustworthy data processing. The fundamental difference from QShield is that they exploit software-based cryptography to realize those security features. Notably, CrytoDB focuses on providing confidentiality for applications running on an untrusted platform with archival data as input; it gives little discussion about access control but just relies on off-the-shelf password-based user management mechanism in DBMSes. Streamforce and PloyStream target at a more challenging real-time stream processing scenario and extend CrytoDB by support multi-user access control. Unfortunately, such improvement introduces significant performance overhead. By comparison, QShield shows its advantage in many aspects. First, it is compatible with both batch and real-time computing paradigms since we opt for a flexible document-oriented data abstraction, which can not only represent tabular data but also can denote data streams. Second, although we only implemented four basic operators in QShield for now, it theoretically is capable of supporting arbitrary operators like ranging and ordering, since SGX supports full CPU arithmetical instructions. Those crypto-based works, however have no such extensibility due to the limitation of cryptography. Third, QShield achieves the two security properties without prohibitive overheads. At present, one limitation of our work is adopting a coarse-grained ACL model; it makes QShield unsuitable to define complex access rules.

\section{Related Work}\label{RW}
Hardware-assisted trusted execution environment (TEE), e.g. Intel SGX, has been utilized in many cloud-based systems. Rearguard \cite{rw_sun_rearguard} leverages the off-the-shelf SGX to enable secure keyword search. As a concurrent work, HardIDX \cite{rw_fuhry} also utilizes it to build secure index for searchable encryption. Opaque \cite{Opaque} and Hermetic \cite{Hermetic} both allow more generic SQL-like queries over encrypted data with SGX. They dedicate themselves to mitigating critical side-channel attacks, but do not offer scalable multi-user access control, nor does it guarantee the correctness of query results. TrustedDB \cite{rw_bajaj} allows data users to execute SQL queries with privacy and under regulatory compliance constraints. It, however, adopts server-hosted, tamper-proof cryptographic coprocessors (SCPUs), which has a much bigger trusted computing base (TCB), instead of SGX to facilitate secure computation in critical query processing stages. Our proposed QShield also exploits SGX to enable secure and efficient SQL-like queries over encrypted data. Assuming a more stronger adversary model, i.e., an untrusted host AS may maliciously invokes enclave interfaces, it achieves scalable access control over multiple data users and guarantees the correctness of query results.

\section{Conclusion}\label{C}
In this paper, we proposed a secure, scalable and efficient SQL-like query framework called QShield for outsourcing data. It utilizes the off-the-shelf hardware-assisted trusted execution environment (TEE), i.e., Intel SGX, to protect the confidentiality and integrity of sensitive data being queried. In order to make QShield capable of enforcing access control over multiple data users in a scalable way, we presented a novel secret sharing mechanism, with which cumbersome authentication through remote attestation per data user is avoided. At the same time, it greatly alleviates attack vectors listed in our threat model. Moreover, we introduced a trust proof mechanism in QShield to guarantee the correctness of query results and further reduce the possibility to build knowledge extractor. We implemented a prototype for QShield and demonstrate that it is feasible in practice. With comprehensive evaluation in terms of both performance and security, we show that QShield achieves fundamental security properties, i.e., confidentiality of crypto key and outsourced data, trustworthy query processing and scalable multi-user access control, while raising no significant performance degradation.

\bibliographystyle{unsrt}
\bibliography{IEEEabrv,myref}

\begin{thebibliography}{10}

\bibitem{CC}
M.~Armbrust, A.~Fox, R.~Griffith, A.~D. Joseph, R.~Katz, A.~Konwinski, G.~Lee,
  D.~Patterson, A.~Rabkin, I.~Stoica, and M.~Zaharia.
\newblock A view of cloud computing.
\newblock {\em Commun. ACM}, 53(4):50--58.

\bibitem{C_Issue}
S.~Hu, C.~Cai, Q.~Wang, C.~Wang, X.~Luo, and K.~Ren.
\newblock Searching an encrypted cloud meets blockchain: A decentralized,
  reliable and fair realization.
\newblock In {\em IEEE INFOCOM 2018 - IEEE Conference on Computer
  Communications}, pages 792--800, April 2018.

\bibitem{cryptdb}
R.~A. Popa, C.~M.~S. Redfield, N.~Zeldovich, and H.~Balakrishnan.
\newblock Cryptdb: Protecting confidentiality with encrypted query processing.
\newblock In {\em Proceedings of the Twenty-Third ACM Symposium on Operating
  Systems Principles}, SOSP '11, pages 85--100, New York, NY, USA, 2011. ACM.

\bibitem{sc_streamforce}
T.~T.~A. Dinh and A.~Datta.
\newblock Streamforce: Outsourcing access control enforcement for stream data
  to the clouds.
\newblock In {\em Proceedings of the 4th ACM Conference on Data and Application
  Security and Privacy}, CODASPY '14, pages 13--24, New York, NY, USA, 2014.
  ACM.

\bibitem{sc_polystream}
C.~Thoma, A.~J. Lee, and A.~Labrinidis.
\newblock Polystream: Cryptographically enforced access controls for outsourced
  data stream processing.
\newblock In {\em Proceedings of the 21st ACM on Symposium on Access Control
  Models and Technologies}, SACMAT '16, pages 227--238, New York, NY, USA,
  2016. ACM.

\bibitem{Ding_PHE}
W.~{Ding}, Z.~{Yan}, and R.~{Deng}.
\newblock Privacy-preserving data processing with flexible access control.
\newblock {\em IEEE Transactions on Dependable and Secure Computing}, pages
  1--1, 2018.

\bibitem{FHE}
C.~Gentry.
\newblock Fully homomorphic encryption using ideal lattices.
\newblock In {\em Proceedings of the Forty-first Annual ACM Symposium on Theory
  of Computing}, STOC '09, pages 169--178, New York, NY, USA, 2009. ACM.

\bibitem{Hermetic}
M.~Xu, A.~Papadimitriou, A.~Haeberlen, and A.~Feldman.
\newblock Hermetic: Privacy-preserving distributed analytics without (most)
  side channels.
\newblock unpublished.

\bibitem{SGX_SA_power}
W.~Wang, G.~Chen, X.~Pan, Y.~Zhang, X.~Wang, V.~Bindschaedler, H.~Tang, and
  C.~A. Gunter.
\newblock Leaky cauldron on the dark land: Understanding memory side-channel
  hazards in sgx.
\newblock In {\em Proceedings of the 2017 ACM SIGSAC Conference on Computer and
  Communications Security}, CCS '17, pages 2421--2434, New York, NY, USA, 2017.
  ACM.

\bibitem{ZeroTrace}
S.~Sasy, S.~Gorbunov, and C.~W.~Fletcher.
\newblock Zerotrace : Oblivious memory primitives from intel sgx.
\newblock In {\em Network and Distributed Systems Security (NDSS) Symposium
  2018}, San Diego, CA, January 2018.

\bibitem{rw_sun_rearguard}
W.~{Sun}, R.~{Zhang}, W.~{Lou}, and Y.~{Thomas Hou}.
\newblock Rearguard: Secure keyword search using trusted hardware.
\newblock In {\em IEEE INFOCOM 2018 - IEEE Conference on Computer
  Communications}, pages 801--809, April 2018.

\bibitem{rw_fuhry}
B.~Fuhry, R.~Bahmani, F.~Brasser, F.~Hahn, F.~Kerschbaum, and A.~R. Sadeghi.
\newblock Hardidx: Practical and secure index with sgx.
\newblock In {\em Data and Applications Security and Privacy XXXI}, pages
  386--408, Cham, 2017. Springer International Publishing.

\bibitem{Opaque}
W.~Zheng, A.~Dave, J.~G. Beekman, R.~A. Popa, J.~E. Gonzalez, and I.~Stoica.
\newblock Opaque: An oblivious and encrypted distributed analytics platform.
\newblock In {\em 14th {USENIX} Symposium on Networked Systems Design and
  Implementation ({NSDI} 17)}, pages 283--298, Boston, MA, 2017. {USENIX}
  Association.

\bibitem{RMF}
Y.~Chen, W.~Sun, N.~Zhang, Q.~Zheng, W.~Lou, and Y.~Hou.
\newblock A secure remote monitoring framework supporting efficient
  fine-grained access control and data processing in iot.
\newblock In {\em Security and Privacy in Communication Networks}, pages 3--21,
  Cham, 2018. Springer International Publishing.

\bibitem{SGX_RANE}
Y.~Swami.
\newblock Intel sgx remote attestation is not sufficient.
\newblock unpublished.

\bibitem{SGX_SA_cache}
J.~G\"{o}tzfried, M.~Eckert, S.~Schinzel, and T.~M\"{u}ller.
\newblock Cache attacks on intel sgx.
\newblock In {\em Proceedings of the 10th European Workshop on Systems
  Security}, EuroSec'17, pages 2:1--2:6, New York, NY, USA, 2017. ACM.

\bibitem{Interface}
J.~Wang, Y.~Cheng, Q.~Li, and Y.~Jiang.
\newblock Interface-based side channel attack against intel sgx.
\newblock unpublished.

\bibitem{VC3}
F.~{Schuster}, M.~{Costa}, C.~{Fournet}, C.~{Gkantsidis}, M.~{Peinado},
  G.~{Mainar-Ruiz}, and M.~{Russinovich}.
\newblock Vc3: Trustworthy data analytics in the cloud using sgx.
\newblock In {\em 2015 IEEE Symposium on Security and Privacy}, pages 38--54,
  May 2015.

\bibitem{Data_Models_1}
R.~{Ranjan}.
\newblock Streaming big data processing in datacenter clouds.
\newblock {\em IEEE Cloud Computing}, 1(1):78--83, May 2014.

\bibitem{Data_Models_2}
D.~{Puthal}.
\newblock Lattice-modelled information flow control of big sensing data streams
  for smart health application.
\newblock {\em IEEE Internet of Things Journal}, pages 1--1, 2019.

\bibitem{SRM}
Y.~Chen, W.~Sun, N.~Zhang, Q.~Zheng, W.~Lou, and Y.~Thomas Hou.
\newblock Towards efficient fine-grained access control and trustworthy data
  processing for remote monitoring services in iot.
\newblock {\em IEEE Transactions on Information Forensics and Security},
  14(7):1830--1842, 2018.

\bibitem{SGX_isolation}
F.~McKeen, I.~Alexandrovich, A.~Berenzon, C.~V. Rozas, H.~Shafi, V.~Shanbhogue,
  and U.~R. Savagaonkar.
\newblock Innovative instructions and software model for isolated execution.
\newblock In {\em Proceedings of the 2Nd International Workshop on Hardware and
  Architectural Support for Security and Privacy}, HASP '13, pages 10:1--10:1,
  New York, NY, USA, 2013. ACM.

\bibitem{SGX_attestation_seal}
I.~Anati, S.~Gueron, S.~Johnson, and V.~Scarlata.
\newblock Innovative technology for cpu based attestation and sealing.
\newblock {\em Workshop on Hardware and Architectural Support for Security and
  Privacy}, 2013.

\bibitem{SGX_explained}
V.~Costan and S.~Devadas.
\newblock Intel sgx explained.
\newblock {\em Cryptology ePrint Archive, Report 2016/086}, 2016.

\bibitem{SGX_SA_branch}
S.~Lee, M.~Shih, P.~Gera, T.~Kim, H.~Kim, and M.~Peinado.
\newblock Inferring fine-grained control flow inside {SGX} enclaves with branch
  shadowing.
\newblock {\em CoRR}, abs/1611.06952, 2016.

\bibitem{SGX_SA_foreshadow}
J.~V. Bulck, M.~Minkin, O.~Weisse, D.~Genkin, B.~Kasikci, F.~Piessens,
  M.~Silberstein, T.~F. Wenisch, Y.~Yarom, and R.~Strackx.
\newblock Foreshadow: Extracting the keys to the intel {SGX} kingdom with
  transient out-of-order execution.
\newblock In {\em 27th {USENIX} Security Symposium ({USENIX} Security 18)},
  page 991{\textendash}1008, Baltimore, MD, 2018. {USENIX} Association.

\bibitem{Code-reuse}
A.~Biondo, M.~Conti, L.~Davi, T.~Frassetto, and A.~Sadeghi.
\newblock The guard{\textquoteright}s dilemma: Efficient code-reuse attacks
  against intel {SGX}.
\newblock In {\em 27th {USENIX} Security Symposium ({USENIX} Security 18)},
  pages 1213--1227, Baltimore, MD, 2018. {USENIX} Association.

\bibitem{Bilinear_Map}
D.~Boneh and M.~Franklin.
\newblock Identity-based encryption from the weil pairing.
\newblock In Joe Kilian, editor, {\em Advances in Cryptology --- CRYPTO 2001},
  pages 213--229, Berlin, Heidelberg, 2001. Springer Berlin Heidelberg.

\bibitem{Goyal_abe}
V.~Goyal, A.~Sahai, O.~Pandey, and B.~Waters.
\newblock Attribute-based encryption for fine-grained access control of
  encrypted data.
\newblock In {\em In Proc. of ACMCCS'06}, pages 89--98, 2006.

\bibitem{Souza_pvss}
R.~D'Souza, D.~Jao, I.~Mironov, and O.~Pandey.
\newblock Publicly verifiable secret sharing for cloud-based key management.
\newblock In Daniel~J. Bernstein and Sanjit Chatterjee, editors, {\em Progress
  in Cryptology -- INDOCRYPT 2011}, pages 290--309, Berlin, Heidelberg, 2011.
  Springer Berlin Heidelberg.

\bibitem{rw_bajaj}
S.~{Bajaj} and R.~{Sion}.
\newblock Trusteddb: A trusted hardware-based database with privacy and data
  confidentiality.
\newblock {\em IEEE Transactions on Knowledge and Data Engineering},
  26(3):752--765, March 2014.

\end{thebibliography}

\end{document}